\documentclass[twocolumn,showpacs,prd]{revtex4}
\usepackage{epsfig}

\begin{document}

\title{MeV-mass dark matter and primordial nucleosynthesis}

\author{Pasquale D.~Serpico and Georg~G.~Raffelt}
\affiliation{Max-Planck-Institut f\"ur
Physik (Werner-Heisenberg-Institut), F\"ohringer Ring 6, 80805
M\"unchen, Germany}

\date{27 May 2004}

\begin{abstract}
  The annihilation of new dark matter candidates with masses $m_X$ in
  the MeV range may account for the galactic positrons that are
  required to explain the 511~keV $\gamma$-ray flux from the galactic
  bulge. We study the impact of MeV-mass thermal relic particles on
  the primordial synthesis of $^2$H, $^4$He, and $^7$Li. If the new
  particles are in thermal equilibrium with neutrinos during the
  nucleosynthesis epoch they increase the helium mass fraction for
  $m_X\alt 10$~MeV and are thus disfavored.  If they couple primarily
  to the electromagnetic plasma they can have the opposite effect of
  lowering both helium and deuterium.  For $m_X=4$--10~MeV they can
  even improve the overall agreement between the predicted and
  observed $^2$H and $^4$He abundances.
\end{abstract}


\maketitle

\section{Introduction}

In a recent series of papers the intriguing possibility was explored
that the cosmic dark matter consists of new elementary particles with
masses in the MeV range~\cite{Boehm:2002yz, Boehm:2003hm,
Boehm:2003bt, Hooper:2003sh, Boehm:2003ha}.  While weakly interacting
massive particles (WIMPs) as dark matter candidates, notably
supersymmetric particles, are usually thought to have masses exceeding
tens of GeV, it is not difficult to come up with viable MeV-mass
candidates such as sterile neutrinos~\cite{Abazajian:2001nj} or
axinos~\cite{Covi:2002vw}. The remarkable aspect of the MeV candidates
studied in Refs.~\cite{Boehm:2002yz, Boehm:2003hm, Boehm:2003bt,
Hooper:2003sh, Boehm:2003ha} is that they are taken to be {\em
thermal\/} relics and thus require a primordial annihilation rate much
larger than given by ordinary weak interactions. Contrary to naive
intuition, such particles are not excluded by any obvious laboratory
measurement or astrophysical argument~\cite{Boehm:2002yz,
Boehm:2003hm, Boehm:2003bt, Hooper:2003sh, Boehm:2003ha, endnote1}.
Quite on the contrary, it was argued that the annihilation of these
dark matter particles in the galactic bulge can produce enough
positrons to explain the 511~keV $\gamma$-ray signature that was
recently confirmed by the INTEGRAL
satellite~\cite{Jean:2003ci,Knodlseder:2003sv} and that seems
difficult to explain with traditional astrophysical sources.
Alternatively, decaying low-mass dark matter particles have also been
proposed as a positron source~\cite{Picciotto:2004rp,Hooper:2004qf}.

In the early universe, weak interactions freeze out at a temperature
of about 1~MeV, just before the epoch of big-bang nucleosynthesis
(BBN). Therefore, MeV-mass particles with larger-than-weak interaction
rates are expected to be in thermal equilibrium at BBN and thus would
add to the primordial energy density and expansion rate. One naively
expects such particles to increase the primordial helium mass fraction
$Y_{\rm p}$, exacerbating the tension between observations and the BBN
prediction. This expectation bears out for new particles coupling
primarily to neutrinos.  However, the new particles proposed in
Refs.~\cite{Boehm:2002yz, Boehm:2003hm, Boehm:2003bt, Hooper:2003sh,
Boehm:2003ha} would primarily interact with the electromagnetic
plasma. It turns out that the BBN effects of these particles are
non-trivial and certainly disfavor masses below about 2~MeV, but
masses in the approximate range 4--10~MeV actually have the opposite
effect of {\em lowering\/} $Y_{\rm p}$ without significantly affecting
deuterium. Therefore, the concordance between the predicted and
observed $^4$He and $^2$H abundances is slightly improved.  This
little-known but intriguing effect was found a long time ago in a
general study of the impact of new particles on
BBN~\cite{Kolb:1986nf}. The main purpose of our work is to re-examine
this effect in the context of the MeV-mass dark matter hypothesis and
with the help of a modern BBN calculation and current observational
data.

In addition to our BBN study presented in Sec.~II we briefly consider
two other possible consequences of MeV-mass dark matter particles. In
Sec.~III we note that the same mechanism proposed to explain the
$\gamma$-ray signature from the galactic bulge should also produce a
diffuse background of low-energy cosmic-ray positrons in the solar
neighborhood.  This argument strengthens the conclusion reached
independently in Refs.~\cite{Boehm:2002yz, Boehm:2003hm, Boehm:2003bt,
Hooper:2003sh, Boehm:2003ha} that an s-wave annihilation cross section
for the reaction $X\bar X\to e^+e^-$ is strongly disfavored.  Further,
in Sec.~IV we consider possible energy-transfer effects caused by such
particles trapped in the Sun or other stars. These effects turn out to
be very small because the trapping efficiency is inhibited by the
smallness of $m_X$. In Sec.~V we summarize our conclusions.

\section{MeV-Mass Particles and Big-Bang Nucleosynthesis}

\subsection{Impact of New Particles}

Thermal relic particles freeze out at a cosmic temperature $T_{\rm F}$
that for weakly interacting particles is about 1~MeV. If the particle
mass is somewhat larger than $T_{\rm F}$ the number density is
suppressed by annihilations so that the relic density is maximal for
$m_X$ of order $T_{\rm F}$, i.e.\ for neutrinos the relic mass density
would be maximal for MeV-range masses and overclose the universe by
about five orders of magnitude~\cite{ket89}.  Therefore, MeV-mass
thermal relics must interact far more strongly than neutrinos so that
they annihilate more efficiently to reduce their density to a level
compatible with the dark-matter abundance.  We briefly summarize the
relationship between the annihilation cross section and the relic
density in Appendix~\ref{app:relicdensity}. The result reveals that
MeV-mass dark matter particles must have been in thermal equilibrium
throughout most of the primordial nucleosynthesis epoch so that the
exact final dark-matter abundance is not relevant for our study.  We
will use the equilibrium assumption in our numerical implementation
that was performed by a modified version of the new BBNCODE recently
developed in Naples and documented in Ref.~\cite{Cuoco:2003cu}.

The impact of our new particles on BBN differs qualitatively and
quantitatively depending on the dominant annihilation and scattering
channels. Before turning to specific cases we discuss how the BBN
equations and inputs are affected.  To this end we first note that the
new particles enter BBN through the second Friedmann equation
$H^2=(8\pi/3)\,G_{\rm N}\,\rho_{\rm tot}$ by their contribution to the
total mass-energy density $\rho_{\rm tot}$.  We consider the
$X$-particles to be in perfect thermal equilibrium and assume that the
number densities of particles and antiparticles are the same for those
cases where $X\not= \bar X$. The number density, pressure, and energy
density contributed by the new particles is thus taken to be
\begin{eqnarray}
n_X&=&\frac{g_X}{2\pi^2}\,T_X^3 \int_x^{\infty}\!\!dy\,
\frac{y\,(y^2-x^2)^{1/2}}{e^y\pm1}\,,\nonumber\\
P_X&=&\frac{g_X}{6\pi^2}\,T_X^4
\int_x^{\infty}\!\!dy\,\frac{(y^2-x^2)^{3/2}}{e^y\pm1}\,,
\nonumber\\
\rho_X&=&\frac{g_X}{2\pi^2}\,T_X^4
\int_x^{\infty}\!\!dy\,\frac{y^2(y^2-x^2)^{1/2}}{e^y\pm1}\,,
\end{eqnarray}
where $x=m_X/T_X$ and the sign $+$ ($-$) refers to fermion (boson)
statistics. $T_X$ is identical with the ambient neutrino temperature
$T_\nu$ for $X$-particles that dominantly couple to neutrinos, while
$T_X=T_\gamma$ for the electromagnetically coupled case.  Since $y\ge
x>0$ and
\begin{equation}
(e^y\pm1)^{-1}=e^{-y}\sum_{n=0}^{\infty}(\mp1)^n e^{-ny}
\end{equation}
we may expand the thermal integrals as
\begin{eqnarray}
&&\kern-1.5em
\int_x^{\infty}\!\!dy\,\frac{y\,(y^2-x^2)^{1/2}}{e^y\pm1}=
x^2\sum_{n=0}^{\infty}(\mp1)^n\frac{K_2\bigl[(1+n)x\bigr]}{1+n}\,,
\nonumber\\
&&\kern-1.5em
\int_x^{\infty}\!\!dy\,\frac{(y^2-x^2)^{3/2}}{e^y\pm1}=
3x^2\sum_{n=0}^{\infty}(\mp1)^n
\frac{K_2\bigl[(1+n)x\bigr]}{(1+n)^2}\,,\nonumber\\
&&\kern-1.5em
\int_x^{\infty}\!\!dy\,\frac{y^2(y^2-x^2)^{1/2}}{e^y\pm1}
\nonumber\\
&&\kern-1.5em{}=
\int_x^{\infty}\!\!dy\,\frac{(y^2-x^2)^{3/2}}{e^y\pm1}
+x^2\int_x^{\infty}dy\frac{(y^2-x^2)^{1/2}}{e^y\pm1}
\nonumber\\
&&\kern-1.5em{}= \sum_{n=0}^{\infty}(\mp1)^n
\left(3x^2\frac{K_2\bigl[(1+n)x\bigr]}{(1+n)^2}+
x^3\frac{K_1\bigl[(1+n)x\bigr]}{1+n}\right)
\,,\nonumber\\
\end{eqnarray}
where $K_l(x)$ is the special Bessel function of order $l$.  In
the numerical code the series are truncated at $n=4$.  The error
of the physical quantities is always $\le 0.18\%$.

A more subtle effect is caused by the conservation of entropy that
implies a modified relation between the neutrino temperature $T_\nu$
and that of the electromagnetic plasma $T\equiv T_\gamma$. The cosmic
entropy density is
\begin{equation}
s=\frac{2\pi^2}{45} g_{s*}T^3=
\frac{2\pi^2}{45}\,T^3\sum_i g_{s*}^i\,,
\end{equation}
where
\begin{eqnarray}
g_{s*}^i(T)&=&\frac{2\pi^2}{45T^3}\frac{\rho_i+P_i}{T_i}\nonumber\\
&=&\left(\frac{T_i}{T}\right)^3
\left(\sum_{\rm bosons} g_i+
\frac{7}{8}\sum_{\rm fermions}g_i\right)\,.
\end{eqnarray}
The second equality applies for relativistic species, i.e.\ in the
absence of warm species where $m\sim T$.  Note that $g_{s*}^i$ is
always considered as a function of the temperature $T$ of the
electromagnetic plasma, not of $T_i$ (a different notation is used in
Ref.~\cite{Kolb:1986nf}).  After neutrino decoupling at $T_{\rm
  D}\approx 2.3$~MeV~\cite{enqv}, the entropy in a comoving volume is
separately conserved for the ``neutrino plasma'' and the
electromagnetic one so that
\begin{equation}\label{tnutgamma}
\left[\frac{g_{s*}^\nu(T_{\rm D})}{g_{s*}^\nu(T)}
\frac{g_{s*}^\gamma(T)}{g_{s*}^\gamma(T_{\rm D})}\right]^{1/3}
=1\,.
\end{equation}
For new particles being only electromagnetically coupled this equation
simplifies to
\begin{equation}
\frac{T}{T_\nu}=
\left[\frac{g_{s*}^{\gamma}(T_{\rm D})}{g_{s*}^{\gamma}(T)}
\right]^{1/3}\,,
\end{equation}
thus giving a higher temperature ratio relative to the standard case.
This simulates the effect of $N_{\rm eff}<3$ thermally excited
neutrino species at BBN and thus a reduced primordial helium
abundance.  The opposite is true for $\nu$-coupled new particles.

The modified $T_\nu(T)$ relation also affects the $n\leftrightarrow p$
weak rates that depend on both $T$ and $T_\nu$ through the phase-space
dependence of the initial and final states of the processes. We
implemented the modification of these rates in a perturbative way
by introducing the small parameter
\begin{equation}
\delta(T)=\frac{T_{\nu}^0(T)-T_\nu(T)}{T_\nu(T)}\,, 
\end{equation}
where $T_{\nu}^0(T)$ is the standard dependence on the electromagnetic
temperature $T$. Typically $\delta$ assumes values of order 0.01 and
is always smaller than about 0.1.  The neutrino temperature enters the
weak rates through Fermi factors of the kind
\begin{equation}
\frac{1}{1+\exp(az_\nu)}\,,
\end{equation}
where $z_\nu=m_e/T_\nu$. Therefore, the additional terms for the rates
can be obtained by integrating the factors
\begin{equation}
\frac{1}{1+\exp[az_\nu^0(1+\delta)]}-\frac{1}{1+\exp(az_\nu^0)}=
\sum_{n=1}^\infty f_n(az_{\nu}^0)\,\delta^n,
\end{equation}
where in our numerical treatment the series was truncated to the third
term. The corrections and the standard rates in the Born approximation
were used to compute numerically the relative changes to the rates.
Therefore, in the limit $\delta\to 0$ one recovers the
$n\leftrightarrow p$ rates including finite mass, QED radiative and 
thermodynamic corrections, while disregarding modifications of these 
subleading effects in the $\delta\not=0$ corrections.  
We finish with six functions $\epsilon_{i}(T)$
\begin{eqnarray}
\frac{\Delta\Gamma_{n\rightarrow p}}{\Gamma_{n\rightarrow p}^0}&=&
\epsilon_{n1}(T)\delta+\epsilon_{n2}(T)\delta^2
+\epsilon_{n3}(T)\delta^3\,,\nonumber\\
\frac{\Delta\Gamma_{p\rightarrow n}}{\Gamma_{p\rightarrow n}^0}&=&
\epsilon_{p1}(T)\delta+\epsilon_{p2}(T)\delta^2
+\epsilon_{p3}(T)\delta^3\,,
\end{eqnarray}
fitting the change in the weak rates with an accuracy better than 5\%.

In principle, one has a single covariant energy conservation equation
for all components of the primordial plasma. For the sake of
simplicity, however, in the previous considerations the ``two fluids
entropy conservation'' was used to obtain the $T_\nu(T)$ relation. We
can now derive the evolution of the thermodynamical quantities by
applying the covariant energy conservation law to one of the two
plasmas, e.g.\ the electromagnetic one, so that the first Friedmann
equation is
\begin{equation}
\frac{dT}{dt}=-3H \left(\rho_{\rm em}+P_{\rm em}\right)
\left(\frac{d\rho_{\rm em}}{dT}\right)^{-1}\,,
\end{equation}
where $H$ depends on $\rho_{\rm tot}$ through the second Friedmann
equation.  If the additional species $X$ couples to the
electromagnetic fluid, the $T$-$t$-relation is further affected by a
modified $(\rho_{\rm em}+P_{\rm em})$ factor, at least until the
scattering freeze-out is reached. This has been roughly estimated to
happen at $T\sim 35$~keV (Appendix~\ref{kinfout}).  In the numerical
code, the $X$-particles were considered to decouple instantaneously
from the electromagnetic component of the plasma for $T\le 35$~keV.
Relaxing this assumption our results remain essentially unchanged
because for such low temperatures nucleosynthesis has almost
completely stopped and the $X$-particles have a negligible impact, at
least for the interesting mass range.

Another input parameter for the BBN calculation is the radiation
density contributed by ordinary neutrinos which we fix to $N_{\rm
  eff}=3$.

Finally, we need the cosmic baryon density.  The best-fit value from
the temperature fluctuations of the cosmic microwave radiation as
measured by the WMAP satellite is $\Omega_{\rm B}h^2=0.024\pm0.001$
\cite{Spergel:2003cb}. Including large-scale structure data from the
2dF galaxy redshift survey shifts this result to $\Omega_{\rm
B}h^2=0.023\pm0.001$, and including Lyman-$\alpha$ data further shifts
it to $0.0226\pm0.0008$ \cite{Spergel:2003cb}. In our study we always
use a fixed value of
\begin{equation}
\Omega_{\rm B}h^2=0.023\,.
\end{equation}
We have checked that for 2$\sigma$ variations of $\Omega_{\rm B}h^2$
our conclusions remain essentially unchanged.  Note that in our BBN
code the final value of $\eta\equiv n_{\rm B}/n_\gamma$ or of
$\Omega_{\rm B}h^2$ is used to work out the one that enters the
initial condition of the problem. We have
\begin{equation}
\eta_{\rm i}=\frac{n_{\rm B}^{\rm i}}{n_\gamma^{\rm i}}
=\frac{n_{\rm B}^{\rm f}}{n_\gamma^{\rm f}}
\frac{n_\gamma^{\rm f}}{n_\gamma^{\rm i}}
\frac{n_{\rm B}^{\rm i}}{n_{\rm B}^{\rm f}}=\eta_{\rm f}
\left(\frac{a_{\rm f}T_{\rm f}}{a_{\rm i}T_{\rm i}}\right)^3
\end{equation}
so that entropy conservation implies the well-known factor 11/4 in the
standard case. In general it is a factor depending on the
$X$-particle properties and was numerically evaluated.

\subsection{Neutrino-Coupled Particles}

\begin{figure}
\vbox{
\hbox to\columnwidth{\hfil\epsfig{file=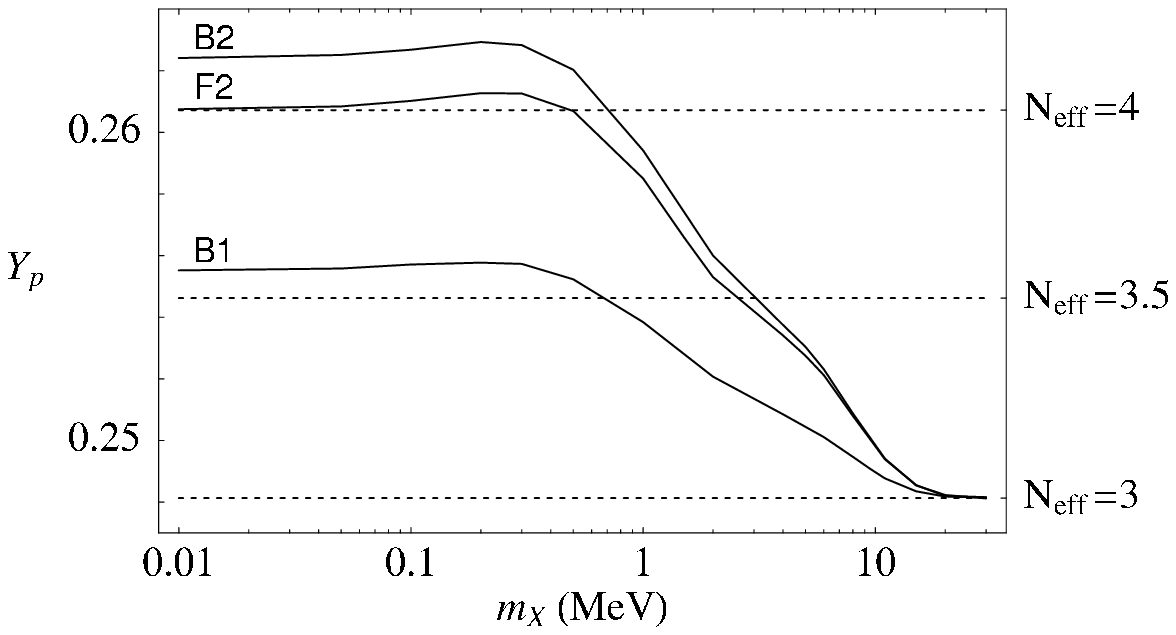,
width=0.92\columnwidth}}
\vspace*{0.8cm}
\hbox to\columnwidth{\hfil\epsfig{file=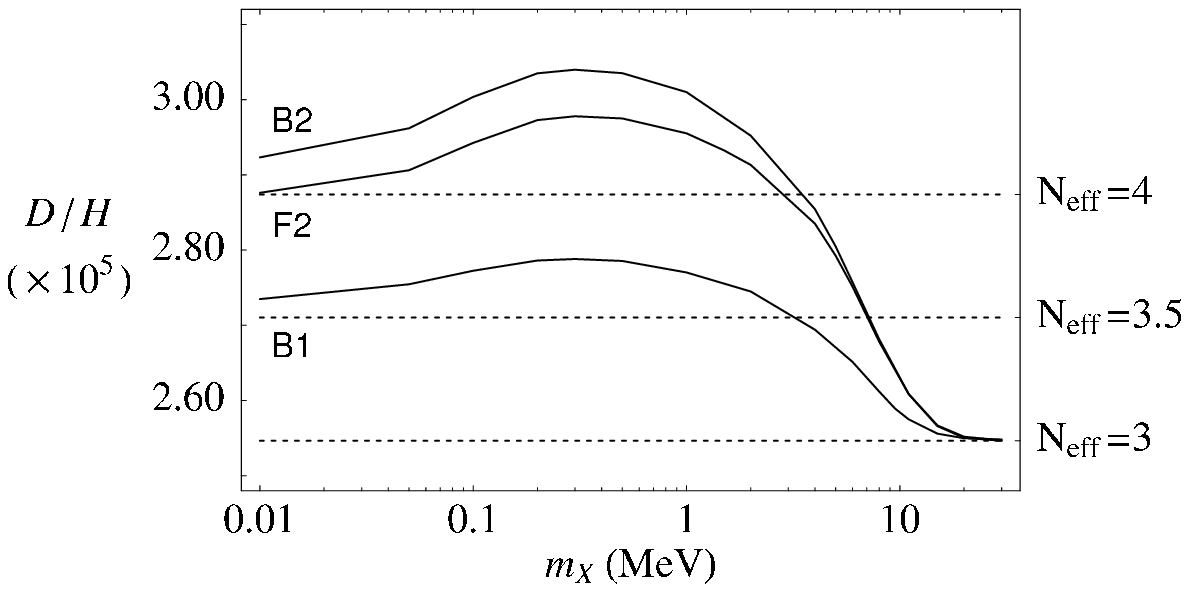,width=0.99\columnwidth}}
\vspace*{0.5cm}
\hbox to\columnwidth{\hfil\epsfig{file=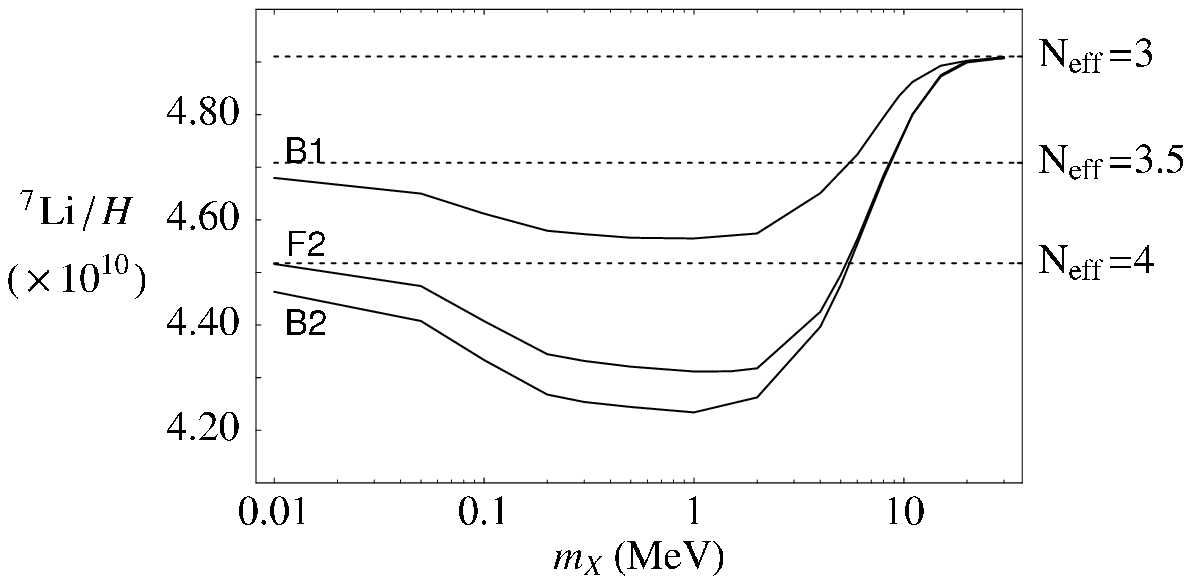,width=\columnwidth}}
\caption{Calculated light-element abundances for $^4$He (top),
$^2$H (middle), and $^7$Li (bottom) for neutrino-coupled new
particles.  The indicated cases B1, B2, and F2 are described in
Table~\ref{tab:cases}. The horizontal lines indicate the effect of
$N_{\rm eff}=3$, 3.5, and 4 relativistic neutrinos.
\label{fig:nucoupled}}
}
\end{figure}

As a first generic type of $X$-particles we consider particles that
annihilate predominantly into neutrinos $X\bar X\leftrightarrow
\nu\bar\nu$. We explicitly study three cases, that of Majorana
fermions with a total of $g_X=2$ inner degrees of freedom (case~F2),
self-conjugate scalar bosons with $g_X=1$ (case~B1), and scalar bosons
with a particle and anti-particle degree of freedom ($g_X=2$,
case~B2). With the ingredients discussed in the previous section we
calculated the abundances for the light elements $^2$H, $^4$He, and
$^7$Li shown in Fig.~\ref{fig:nucoupled} as a function of the new
particle mass $m_X$.  For $m_X\agt20$~MeV we recover the standard BBN
predictions. For very small masses $m_X\to 0$ these particles freeze
out relativistically and their effect on BBN is exactly that of
$\Delta N_{\rm eff}=4/7$, 1, or 8/7 additional relativistic neutrinos
for the three cases B1, F2 and B2, as summarized in
Table~\ref{tab:cases}.

\begin{table}[b]
\caption{\label{tab:cases}Cases for new particles}
\begin{ruledtabular}
\begin{tabular}{llllllll}
\multicolumn{2}{l}{Case}&$g_X$&$\Delta N_{\rm eff}$&
$\eta_{\rm i}/\eta_{\rm f}$\\
B1&Self-conjugate scalar boson&1     &4/7 & 3.25\\
B2&Scalar boson $X\not=\bar X$ &2     &8/7 & 3.75\\
F2&Majorana fermion    &2     &1 & 3.65\\
\end{tabular}
\end{ruledtabular}
\end{table}

The BBN effect of the new particles is dominated by their contribution
to the primordial energy density and thus similar to an additional
neutrino species. However, it is worthwhile to note the extra effect
for intermediate $m_X$ relative to the asymptotic case $m_X\to 0$. It
is caused by the modified $T_\nu/T$ ratio previously described.

Our results essentially agree, both qualitatively and quantitatively,
with those of Ref.~\cite{Kolb:1986nf}, except for lithium. The
difference is explained by our value of $\Omega_{\rm B}h^2$ where the
dependence of $^7$Li on $N_{\rm eff}$ is opposite from the situation
in Ref.~\cite{Kolb:1986nf}.  In our case this nuclide is essentially
produced through the channel ${}^4{\rm He}({}^3{\rm
He},\gamma){}^7{\rm Be}(e^-,\nu_e){}^7{\rm Li}$ while at the lower
value of $\Omega_{\rm B}h^2$ used in Ref.~\cite{Kolb:1986nf} the
direct channel ${}^4{\rm He}({}^3{\rm H},\gamma) {}^7{\rm Li}$
dominates.

Our theoretical predictions can be compared with the measured
primordial abundances summarized in Table~\ref{tab:abundances}.  For
helium, the standard BBN prediction significantly exceeds the most
recent measured value~\cite{Izotov:2003xn}, and this discrepancy is
even worse for other helium determinations that are lower (for a
review see, e.g.\ Ref.~\cite{Hagiwara:fs}).  Therefore, while
different observations of the primordial $^4$He abundance disagree on
its exact value, a tension with the BBN prediction always exists.  Our
new particles exacerbate this discrepancy for $m_X\alt 10$~MeV and are
thus disfavored or even excluded.

The deuterium abundance extracted from the QSO systems agrees
perfectly with the BBN prediction, although it could be affected by
possibly underestimated systematic errors. In any event, the new
particles change the prediction only within the $1\sigma$
observational range so that deuterium adds little new information on
the viability of the new particles.

A clear interpretation of the Spite plateau in the lithium data from
metal-poor halo stars is still lacking so that it is not clear how to
compare the lithium observations (average value given in
Table~\ref{tab:abundances}) with the theoretical prediction that is
roughly a factor 2--3 larger. Both standard and non-standard physics
explanations of this discrepancy have been invoked, e.g.\
Ref.~\cite{Jedamzik:2004er} and references. Therefore, a meaningful
comparison of our non-standard BBN prediction for lithium with
observations is difficult. We show these results primarily for the
purpose of illustration.

\begin{table}
\caption{\label{tab:abundances}Measured primordial light-element
abundances}
\begin{ruledtabular}
\begin{tabular}{llll}
\multicolumn{2}{l}{Element}&Abundance&Reference\\
Helium\footnote{For other determinations and a short review, see also~\cite{Hagiwara:fs}. }   &$^4$He&$Y_{\rm p}=0.2421\pm0.0021$&\cite{Izotov:2003xn}\\
Deuterium&$^2$H &${\rm D/H}=2.78^{+0.44}_{-0.38}\times10^{-5}$&
\cite{Kirkman:2003uv}\\
Lithium  &$^7$Li&$^7{\rm Li/H}=(1.70\pm0.17)\times10^{-10}$&
Refs.\ in \cite{Cuoco:2003cu}\\
\end{tabular}
\end{ruledtabular}
\end{table}

\subsection{Electromagnetic Couplings}

We next turn to new particles that do not interact with neutrinos but
remain in perfect thermal equilibrium with the electromagnetic plasma
throughout the BBN epoch by virtue of processes such as $X\bar
X\leftrightarrow e^+e^-$ and $X+e\leftrightarrow X+ e$.  The BBN
impact of these particles is more subtle and can be opposite to the
neutrino-coupled case. Moreover, this is the case relevant as a source
for galactic positrons~\cite{Boehm:2002yz, Boehm:2003hm, Boehm:2003bt,
Hooper:2003sh, Boehm:2003ha}.  For particles of this sort, there is a
tension between the required primordial annihilation cross section to
achieve the correct dark-matter density and the one in the galaxy to
avoid overproducing positrons. The solution prefered in
Refs.~\cite{Boehm:2002yz, Boehm:2003hm, Boehm:2003bt, Hooper:2003sh,
Boehm:2003ha} is that of a predominant p-wave annihilation channel
that will suppress the annihilation rate in the galaxy relative to
that in the early universe. In particular, a specific model for a new
boson was constructed where particles and anti-particles are not
identical, i.e.\ our case B2.  Of course, the detailed form of the
cross section is not relevant for our work because we assume that the
new particles are in perfect equilibrium with the electromagnetic
plasma.

\begin{figure}
\vbox{
\hbox to\columnwidth{\hfil\epsfig{file=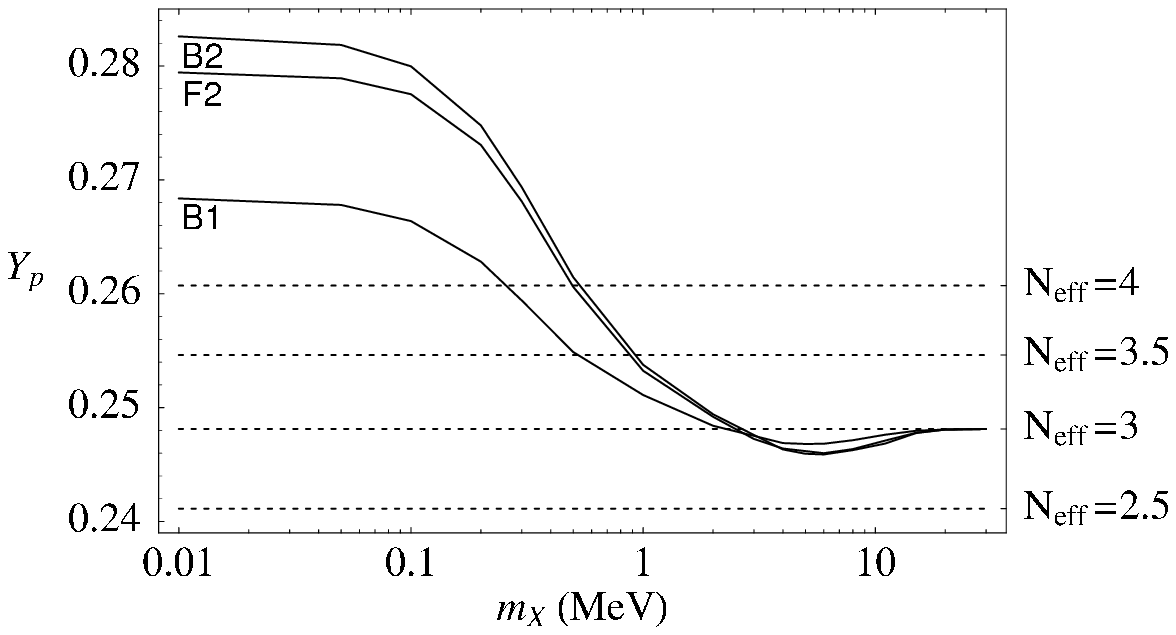,
width=0.95\columnwidth}}
\vspace*{0.4cm}
\hbox to\columnwidth{\hfil\epsfig{file=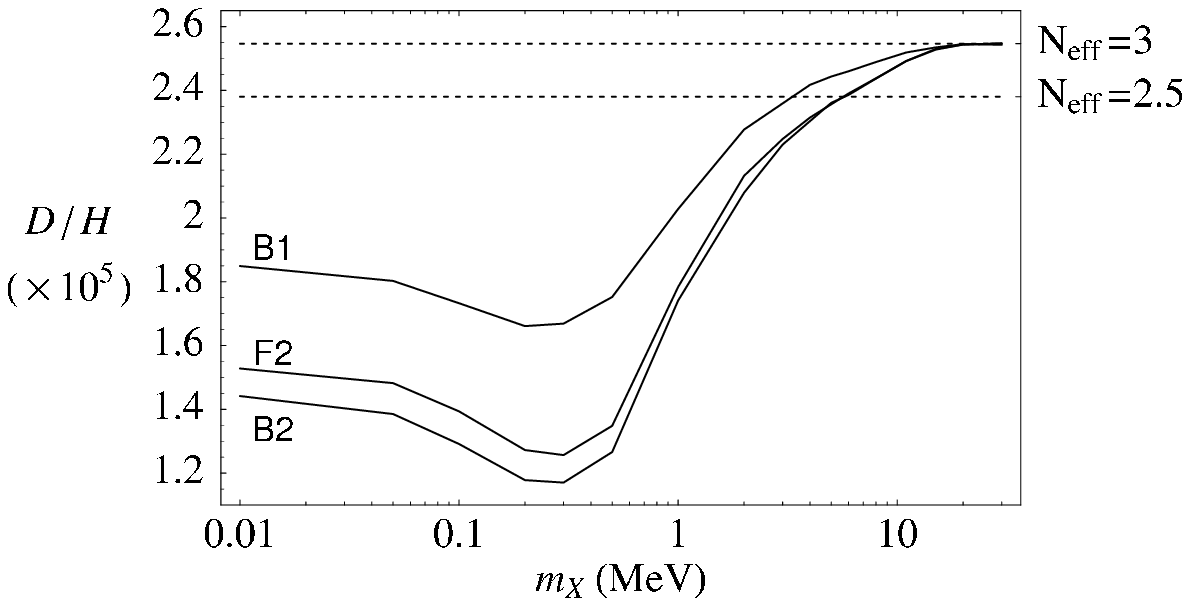,
width=\columnwidth}}
\vspace*{0.5cm}
\hbox to\columnwidth{\hfil\epsfig{file=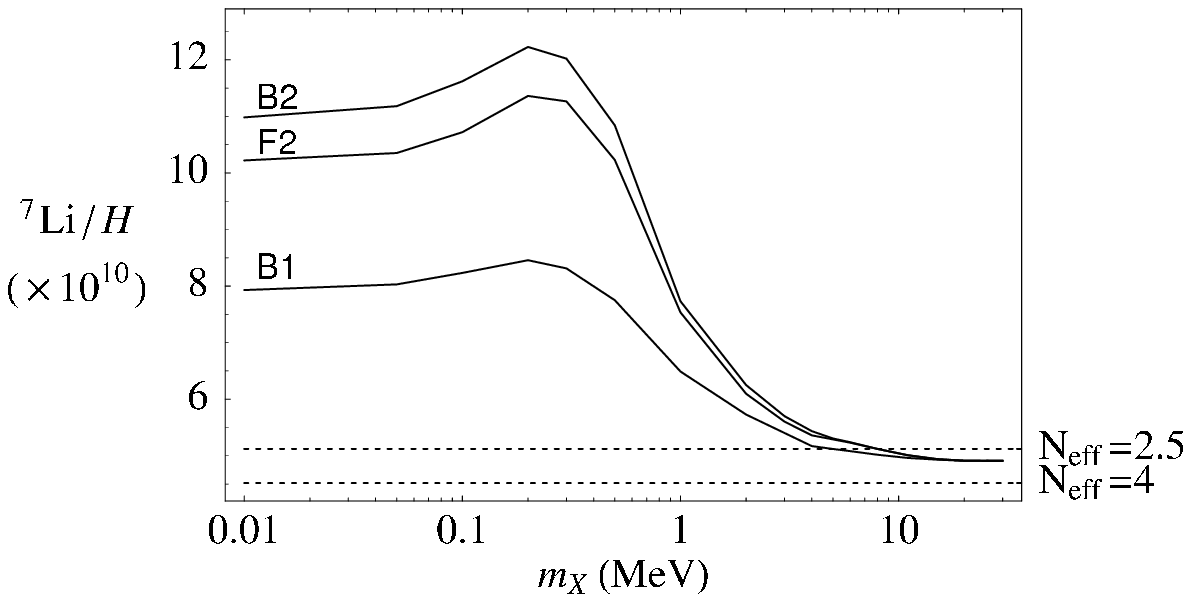,width=\columnwidth}}
\caption{Calculated light-element abundances for $^4$He (top), $^2$H
(middle), and $^7$Li (bottom) for electromagnetically coupled new
particles.  The indicated cases B1, B2, and F2 are described in
Table~\ref{tab:cases}.  The horizontal dotted lines indicate the
standard predictions for the indicated values of the effective number
of relativistic neutrino species, $N_{\rm eff}$.
\label{fig:emcoupled}}
}
\end{figure}

\begin{figure}
\epsfig{file=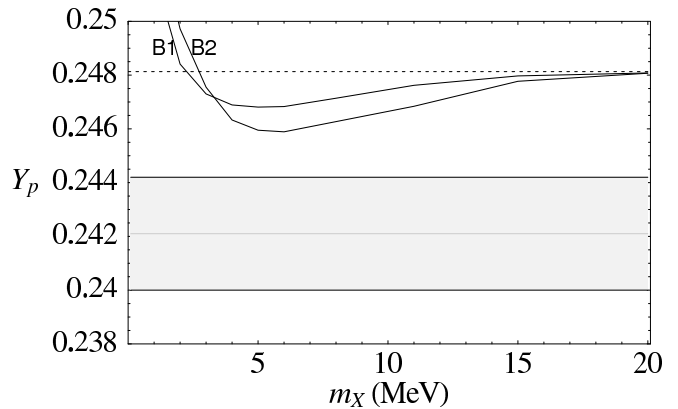,width=\columnwidth}
\caption{$^4$He abundance as in Fig.~\ref{fig:emcoupled}, here on a
linear scale for $m_X$.  The horizontal dotted line indicates the
standard prediction for $N_{\rm eff}=3$. The gray band is the
1$\sigma$ observational range for $Y_{\rm p}$ according to
Ref.~\cite{Izotov:2003xn}.
\label{fig:He4obs}}
\end{figure}

The calculated light-element abundances as functions of $m_X$ are
shown in Fig.~\ref{fig:emcoupled}.  For $m_X\alt 2$~MeV the abundances
are all shifted away from the observed values.  However, for $m_X\agt
2$~MeV the ``entropy effect'' works in the direction of lowering
$Y_{\rm p}$, by up to $\Delta Y_{\rm p}\approx -0.002$ for the B2
case, without significantly affecting the $^2$H and $^7$Li
predictions.  Therefore, the concordance with the $^4$He observations
is improved. This point is illustrated more directly by
Fig.~\ref{fig:He4obs} where we compare the new $^4$He predictions for
the cases B1 and B2 with the standard $N_{\rm eff}=3$ case and with
the recent observational determination~\cite{Izotov:2003xn} for which
the 1$\sigma$ error band is shown.  $Y_{\rm p}$ would benefit from
this effect up to $m_X\alt 15$~MeV, even if the values $m_X\alt
10$~MeV are prefered. Our results again agree qualitatively with those
of Ref.~\cite{Kolb:1986nf}.

For low masses, the value of $\eta_{\rm i}$ needed to match the WMAP
finding for $\eta_{\rm f}$ is significantly increased relative to the
standard factor 2.75, as shown in the last column of
Table~\ref{tab:cases}. This explains physically the huge decrease
(increase) in the $^2$H ($^7$Li) yield that does not strongly depend
on $N_{\rm eff}$. Quite on the contrary, as $Y_{\rm p}$ depends only
logarithmically on $\eta$, its change is essentially dominated by the
addition of extra degrees of freedom to the electromagnetic
plasma. The $X$-particles are now hotter than in the neutrino case and
thus provoke a bigger effect.  For $m_X\agt 20$~MeV one recovers
the standard predictions.

We have always neglected the role of dark matter residual
annihilations during the freeze-out epoch.  A detailed treatment of
such effects is beyond the scope of our paper. However, an approximate
study of this phenomenon (Appendix~\ref{decouplentr}) indicates that
this late-time entropy generation effect is sub-dominant.  Its effect
goes in the direction of a further marginal reduction of $Y_{\rm p}$
and a small increase in the deuterium yield.

We have also neglected the possible photo-dissociation of $^2$H and
$^7$Li induced by late dark matter annihilations. As shown in
Appendix~\ref{Ddissoc}, this effect is marginal unless, perhaps,
if the dark matter particles couple directly with photons.

\subsection{Particles Coupled both to the Electromagnetic Plasma
and to Neutrinos}

It may also be that the new particles interact strongly enough with
both charged leptons and neutrinos to keep the two fluids in
equilibrium beyond the usual decoupling epoch, a situation that was
not previously treated.  In this case $T_\gamma=T_\nu$ is maintained
longer and perhaps throughout the nucleosynthesis epoch, depending on
the new particle properties.  This effect would be present even for
very high $m_X$ if one introduces an additional direct coupling of the
neutrinos to the charged leptons through an extra gauge boson as in
the prefered $U$-boson model discussed in Refs.~\cite{Boehm:2002yz,
Boehm:2003hm, Boehm:2003bt, Hooper:2003sh, Boehm:2003ha}.  Of course,
in this case laboratory data, e.g.~from $\nu$-$e$
scattering~\cite{LAMPF,LSND}, provide strong limits so that this
situation may be rather unphysical.

\begin{figure}
\vbox{
\hbox to\columnwidth{\hfil\epsfig{file=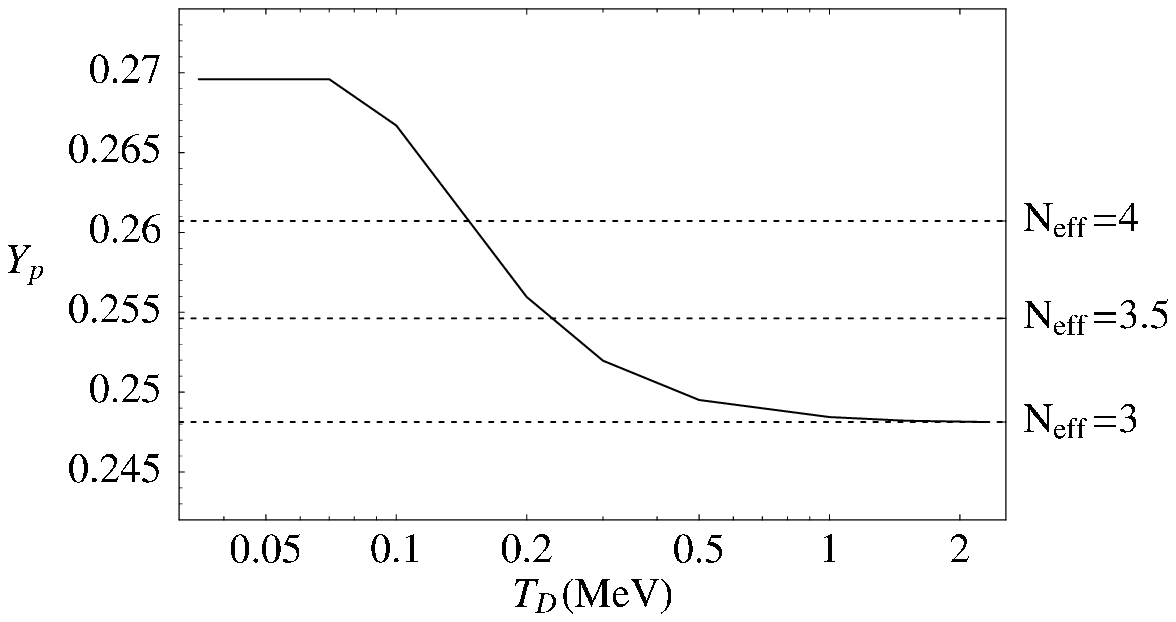,width=0.99\columnwidth}}
\vspace*{0.9cm}
\hbox to\columnwidth{\hfil\epsfig{file=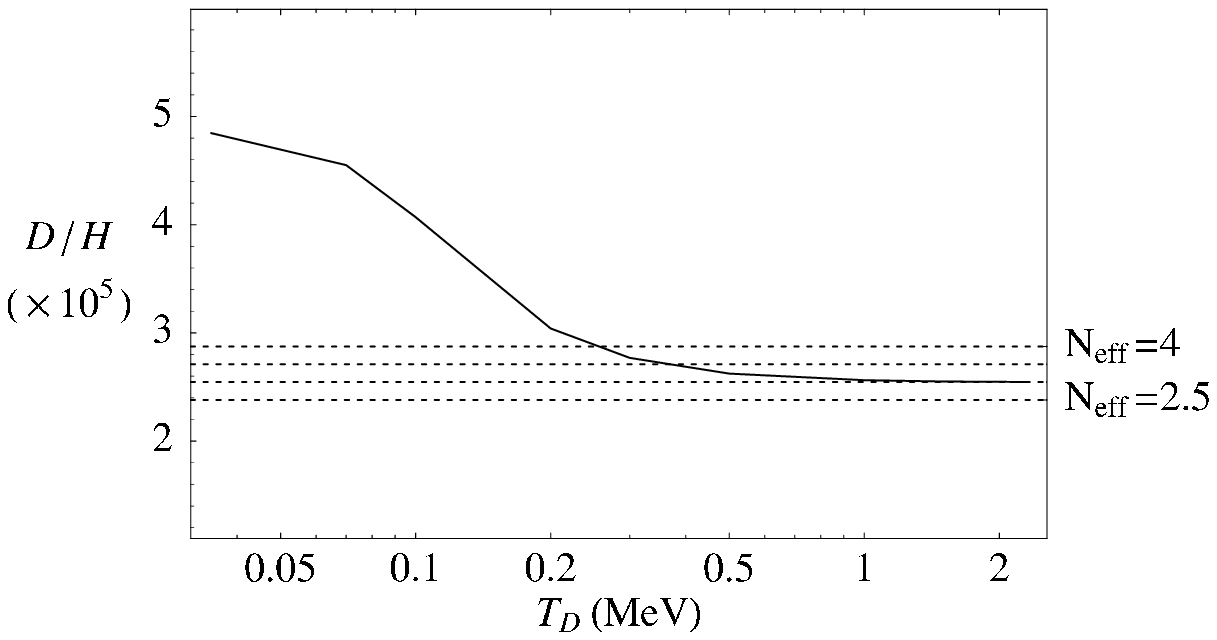,width=0.99\columnwidth}}
\vspace*{0.6cm}
\hbox to\columnwidth{\hfil\epsfig{file=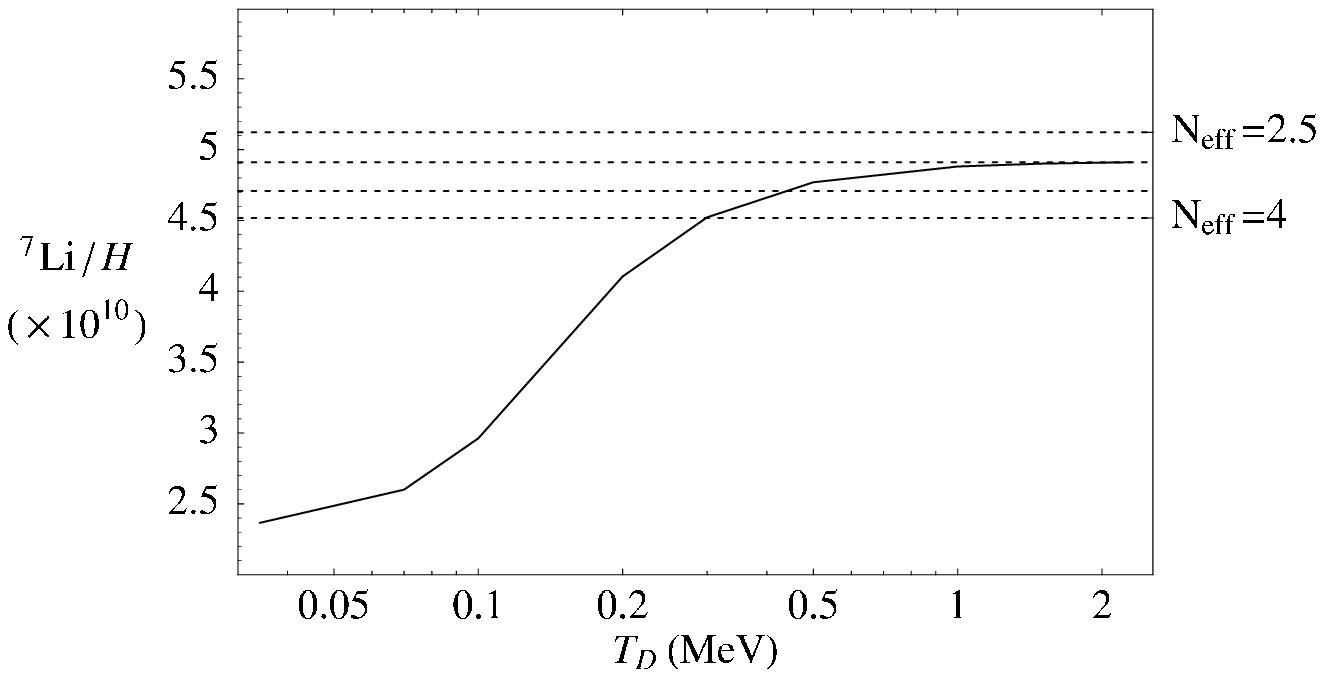,width=\columnwidth}}
\caption{
\label{fig:bothcoupled}Calculated light-element abundances for 
$^4$He (top), $^2$H (middle), and $^7$Li (bottom) if the neutrino and
electromagnetic plasmas are thermally coupled by a new interaction
channel down to a temperature $T_{\rm D}$.  The horizontal lines
indicate the standard predictions for the indicated values of the
effective number of relativistic neutrino species, $N_{\rm eff}$.}  }
\end{figure}

In Fig.~\ref{fig:bothcoupled} we show the light-element abundances as
a function of the assumed neutrino decoupling temperature $T_{\rm D}$.
If it is lowered to values $T_D\alt m_e$, the resulting modification
of the light-element abundances is quite extreme and strictly
excluded. For $^2$H and $^7$Li the ``eta-effect'' dominates, while for
$^4$He the change in $T_\nu/T$ and that of $\rho_\nu$ play a major
role.

If the neutrinos and electromagnetic fluids are only indirectly
coupled by their interactions with the new $X$-particles, the limiting
condition to prolong the $\nu$-equilibrium with the photon bath is the
number density factor of the dark matter, $n_X$, in addition to the
interaction cross sections.  Qualitatively, we expect that for
$m_X\alt m_e$ the combined effects conspire to hugely increase the
$^4$He yield, while some compensation could appear in the $^2$H and
$^7$Li behavior.  Obviously, for $m_X\agt20$~MeV one should
recover the standard predictions, unless a direct $\nu$-$e$ coupling
exists.  A detailed kinetic study of this case would require a
concrete particle physics model and is beyond the scope of our work.

\section{Low-energy cosmic ray positrons}

The proposed MeV-mass dark matter particles discussed in
Refs.~\cite{Boehm:2002yz, Boehm:2003hm, Boehm:2003bt, Hooper:2003sh,
Boehm:2003ha} are supposed to annihilate in the galactic bulge and
produce a flux of low-energy cosmic-ray positrons that can explain the
observed 511~keV $\gamma$-ray signature.  For such $X$-particles we
also expect a flux of low-energy positrons at Earth from local
dark-matter annihilation.  Such a signature was already proposed for
the detection of the traditional GeV--TeV mass range of dark matter
particle candidates~\cite{Tylka:xj,Turner:1989kg}.  For the case of
annihilating MeV-mass dark matter particles we expect a much larger
positron flux and therefore we study if additional constraints arise
from the low-energy cosmic-ray positrons in the solar neighborhood.

To this end we derive the expected positron flux from $X$-particle
annihilation.  For a stationary situation the continuity equation for
the cosmic-ray positrons is
\begin{equation}\label{conteq}
q(E)-\frac{n(E)}{\tau(E)}+\frac{d [b(E)n(E)]}{dE}=0\,,
\end{equation}
where $n(E)$ is the differential positron density, $q(E)$ the
injection spectrum, $\tau(E)$ an effective containment time, and
$b(E)\equiv -dE/dt$ the energy-loss function.
This equation is solved by
\begin{equation}\label{solution}
n(E)=\frac{1}{b(E)}\int_E^\infty ds\, q(s)
\exp\left(-\int_E^s\frac{dy}{\tau(y)b(y)}\right)\,.
\end{equation}
The expected positron flux at the top of the atmosphere, neglecting
solar modulation effects, is $j_{e^+}=n(E)/4\pi$, where we have
assumed relativistic velocities.

In the relevant energy range and for an essentially neutral
environment, ionization and bremsstrahlung are the dominant
energy-loss mechanisms so that we have~\cite{schlieckbook}
\begin{eqnarray}
b(E)&\approx&36.18\,m_e\sigma_{\rm T} n_{\rm H}
\left[1+0.146\,\ln\left(\frac{E}{m_e}\right)
+\frac{E}{709\,m_e}\right]\nonumber\\
&=&3.69\times 10^{-13}~{\rm MeV~s^{-1}}\,
n_{\rm H}\, [\ldots]\,.
\end{eqnarray}
Here, $\sigma_{\rm T}$ is the Thomson cross section and $n_{\rm H}$
the hydrogen density that dominates the interstellar medium.  For the
relevant energies, $b(E)$ is a slowly varying function.

It is easy to estimate that low-energy positrons do not travel far
before annihilating so that the containment time $\tau(E)$ is
identical with the annihilation time scale.  The latter
can be written in the form~\cite{Svensson:hz}
\begin{equation}
\tau(E) =
\frac{1.33\times 10^{14}~\rm s}{f(E)\,n_{\rm H}}\,,
\end{equation}
where $f(E)=0.02$--1 in the energy range of interest.

Finally we need the positron injection spectrum from dark-matter
annihilation,
\begin{equation}
q(E)=n_X^2\langle\sigma_a v\rangle\delta(1-E/m_X)\,,
\end{equation}
where we have assumed that the $X$-particles are different from their
antiparticles and that $n_X=n_{\bar X}$. 

Equation~(\ref{solution}) cannot be expressed in closed form, but for
our purposes an analytical approximation is accurate enough.  The
factor $b\tau$ does not depend on $n_{\rm H}$ and is found to be
\begin{equation}\label{kappadef}
b \tau=\hbox{50--5000~MeV}\,,
\end{equation}
where the range reflects the monotonic energy dependence.  Ignoring
this energy-dependence allows us to write the solution of
Eq.~(\ref{solution}) for $E\leq m_X$ as
\begin{equation}
n(E)\approx 
\frac{n_X^2 \langle\sigma_a v\rangle}{b(E)}\,
\exp\left(-\frac{m_X-E}{b\tau}\right)\,.
\end{equation}
For the interesting range of $m_X$ and $E$ the exponential factor is
always close to 1. Physically this represents the fact that positrons
produced at energy $E=m_X$ will be lost from this ``energy bin''
primarily by down-scattering, not by annihilation.

In order to predict the positron flux we use $n_{\rm H}=1~\rm
cm^{-3}$. For the local dark matter density we use the canonical value
$\rho_{\rm DM}= 300~{\rm MeV~cm^{-3}}=m_X(n_X+n_{\bar X})$.  For the
annihilation cross section we first consider an s-wave model with
$\langle\sigma_a v\rangle=\sigma_0$ where $\sigma_0$ is fixed by the
early-universe freeze-out calculation
(Appendix~\ref{app:relicdensity}). We compare the flux prediction with
the best 95\% CL upper limits in the 20--90 MeV range that are given
in Fig.~4 of Ref.~\cite{hp76}.  At 20~MeV, the flux limit is
approximately $1.2\times 10^{-5}~\rm cm^{-2}~s^{-1}~sr^{-1}~MeV^{-1}$,
for 30~MeV it is $8.5\times 10^{-6}$ in these units, for 50~MeV it is
$2.5\times 10^{-6}$, and for 90~MeV it is $9\times 10^{-7}$.

We find flux predictions exceeding these limits by factors
${}\agt500$. This implies that the annihilation cross section required
in the early universe for reducing the particle density to the
dark-matter level produces an unacceptably large local positron flux
and is thus excluded for $m_X>20$~MeV. For smaller $m_X$ we have no
constraint. For a p-wave channel we do not obtain any limits because
$\langle\sigma_a v^2\rangle$ is strongly suppressed in the galaxy
because of the small dark-matter velocities $v$ of order~$10^{-3}$.

Therefore, we confirm the conclusion of Refs.~\cite{Boehm:2002yz,
Boehm:2003hm, Boehm:2003bt, Hooper:2003sh, Boehm:2003ha} that an
s-wave annihilation channel $X\bar X\to e^+e^-$ is not acceptable for
thermal relics. However, our argument does not depend on the uncertain
dark-matter profile of the galactic bulge.

\section{Energy Transfer in Stars}

New MeV-mass particles could play an important role in stars. For
example, they would be thermally produced in the collapsed core of a
supernova and could contribute to the energy loss or the transfer of
energy in these systems~\cite{Raffelt:1999tx}. However, the
$X$-particles discussed here have stronger-than-weak interactions,
implying that in a supernova these effects would be small compared to
those of ordinary neutrinos. Therefore, even though the new particles
would be thermally excited in a supernova core, there are no obvious
observational consequences.

In ordinary stars, and especially in our Sun, dark-matter particles
will be trapped and contribute to the transfer of energy in
potentially observable ways~\cite{sp85}. In the following we
investigate if MeV-mass particles could be relevant in this context.
The result is that for MeV-range masses the evaporation time is very
short so that the steady-state abundance of $X$-particles in the Sun
is too small to be important.

A simple estimate of the energy conduction by the new particles can be
worked out in a one-zone model of the Sun~\cite{sp85}.  One assumes
that the dark-matter stationary distribution in the Sun is globally
Maxwellian at a uniform temperature $T_X$, with its maximum density
found at a scale radius $r_X$. Assuming that $T_X$ is identical with
the temperature at the solar center, and taking $r_X$ to be of order
the solar radius, the luminosity carried by the new particles is of
order
\begin{equation}\label{lumx}
\frac{L_X}{L_\odot}\sim 
\frac{10^{-42}\,N_X}{(\sigma_s/{\rm pb})\sqrt{m_X/{\rm MeV}}}\,,
\end{equation}
where $N_X$ is the total number of $X$-particles trapped in the Sun
and $\sigma_s$ is the scattering cross section on electrons, taken to
be comparable to the annihilation cross section as described in
Appendix~\ref{kinfout}.  For particles even more weakly interacting,
i.e.\ for $\sigma_s\ll{}$few~pb, one enters the Knudsen regime where
the effect of energy transfer is much
smaller~\cite{Gould:1989ez,Gould:hm}.

The steady-state number of dark-matter particles collected by the Sun
arises from an equilibrium between capture and evaporation, i.e.\
$N_X=A/P_e$ with $A$ the number of particles captured per unit time
and $P_e$ the escape probability per unit time.  We estimate the
capture rate to be~\cite{Press:ug,Gould:1987ir}
\begin{equation}
A\sim 4.34\,\frac{n_XG_{\rm N}M_{\odot}R_{\odot}}{v_{\rm gal}}
\end{equation}
where $v_{\rm gal}\approx300~\rm km~s^{-1}$ is the mean square
velocity of the galactic dark matter near the orbit of the Sun and
$n_X$ its number density.

The escape probability $P_e$ is physically the ratio of the fraction
of particles in the ``escape-tail'' of their distribution to the
typical time needed to repopulate it. A simple estimate is~\cite{sp85}
\begin{equation}
P_e\sim \frac{(v_f \delta v)^{3/2}}{G_{\rm N} M_{\odot}}
\sqrt{\frac{m_X v_f^2}{2T_X}}
\exp\left(-\frac{m_X v_f^2}{2T_X}\right),
\end{equation}
where $v_f$ is the escape velocity from a typical position in the Sun.
Near the solar center one has $v_f\sim(15.8\,T_X/m_p)^{1/2}$
\cite{Press:ug}. Further, $\delta v\sim v_s m_s/m_X$ where $m_s$ and
$v_s$ refer to the scattered particles, electrons in our case, for
which we assume a thermal velocity distribution so that
$v_s=(3T_s/m_s)^{1/2}$ and we assume $T_s=T_X$.

Based on these simple estimates and using a simplified solar model we
find $N_{X}\sim 10^{36}\,({\rm pb}/\sigma_s)$.  Comparing this result
with Eq.~(\ref{lumx}) we conclude that the effect of the new
particles is too small to be significant for the Sun.  Of course, our
estimate is rather crude considering that the escape lifetime of the
dark matter particles is comparable to their orbital period.

\section{Summary and Conclusions}

We have analyzed some astrophysical and cosmological consequences of
the intriguing possibility that the cosmic dark matter consists of
MeV-mass particles~\cite{Boehm:2002yz, Boehm:2003hm, Boehm:2003bt,
Hooper:2003sh, Boehm:2003ha}.  These particles are assumed to be
thermal relics and thus have interaction cross sections that are
larger than weak.

Such particles do not have any apparent consequences for stellar
evolution. In supernovae, their interaction is too strong so that
neutrinos continue to play the leading role for energy loss and energy
transfer.  In the Sun, the particle mass is so small that evaporation
prevents the trapped steady-state population from growing large enough
for a significant contribution to energy conduction.

We have derived new constraints coming from the low-energy positron
component of cosmic rays. An s-wave annihilation cross section for
$X\bar X\to e^+e^-$ as large as implied by the early-universe
freeze-out calculation causes an excessive positron flux for
$m_X=20$--90~MeV where experimental upper limits are
available. Therefore, only p-wave annihilation is compatible with
these constraints. These conclusions agree with those reached in
Refs.~\cite{Boehm:2002yz, Boehm:2003hm, Boehm:2003bt, Hooper:2003sh,
Boehm:2003ha}, but in our case they do not depend on the assumed
dark-matter profile of the galactic bulge.

The main subject of our work, however, was the impact of MeV-mass
particles on big-bang nucleosynthesis (BBN). Significant modifications
arise only for $m_X\alt20$~MeV.  The effects found are largely
independent of the energy dependence of the annihilation cross section
and even of its exact value, provided that it is high enough to take
the new particles to equilibrium with the neutrinos or the
electromagnetic plasma.

For the neutrino-coupled case, the impact of the new particles is
comparable to that of additional neutrino species.  Notably, the
primordial helium abundance is increased, exacerbating the discrepancy
between predictions and observations.  Therefore, such $X$-particles
are disfavored by BBN for masses up to about 10~MeV.

For the electromagnetically coupled case, the BBN concordance would be
severely disturbed for $m_X\alt2$~MeV.  However, there is a region
$m_X=4$--10~MeV where the primordial helium mass fraction $Y_{\rm p}$
is actually reduced relative to the standard case while the predicted
$^2$H remains compatible with observations.  This non-trivial
phenomenon is a consequence of the ``entropy effect'' discussed in the
paper.  While this effect was already found in
Ref.~\cite{Kolb:1986nf}, it now assumes greater importance because it
slightly improves the discrepancy between the BBN predictions for
$Y_{\rm p}$ and its observed value.

In summary, the MeV-mass dark matter particles proposed in
Refs.~\cite{Boehm:2002yz, Boehm:2003hm, Boehm:2003bt, Hooper:2003sh,
Boehm:2003ha} as a source for positrons in the galactic bulge are not
incompatible with BBN, provided the particle mass exceeds a few
MeV. On the contrary, in the mass range $m_X=4$--10~MeV these
particles slightly improve the concordance between BBN calculations
and the observed helium abundance. It is quite fascinating that such
exotic particles, far from being excluded, seem to have several
beneficial consequences in astrophysics and cosmology.


\begin{acknowledgments}
  
  In Munich, this work was supported in part by the Deut\-sche
  For\-schungs\-ge\-mein\-schaft under grant SFB 375 and the ESF
  network Neutrino Astrophysics. We thank D.~Semikoz, M.~Kachelriess
  and the members of the Naples Astroparticle Group for useful
  discussions and comments.

\end{acknowledgments}

\appendix

\section{Relic Density and Annihilation Cross Section}
\label{app:relicdensity}

Once the mass $m_X$ and the spin multiplicity $g_X$ of a thermal relic
are fixed, the requirement that it is the dark matter uniquely
determines the annihilation freeze-out temperature $T_{\rm F}$ and the
annihilation reaction rate at this temperature $\langle\sigma_a
v\rangle_{T_{\rm F}}$. The relic mass density is
\begin{equation}
\Omega_{X}h^2=
\frac{8 \pi h^2 m_X n_X}{3 m_{\rm Pl}^2H^2}=
\frac{8 \pi m_X s}{3(m_{\rm Pl}H/h)^2}\left(\frac{n_X}{s}\right)\,,
\end{equation}
where $H/h=100\rm~km~s^{-1}~Mpc^{-1}$, $m_{\rm Pl}=1.22\times
10^{22}$~MeV is the Planck mass, and $s$ is the entropy density.  Note
that for particles that are not self-conjugate $\Omega_X$ only
includes the particles so that $\Omega_{\rm DM}=\Omega_X+\Omega_{\bar
X}$.

Neglecting entropy-producing phenomena, the entropy per comoving
volume remains constant so that
\begin{equation}
\left(\frac{n_X}{s}\right)_0=\left(\frac{n_X}{s}\right)_{T_{\rm F}}
\end{equation}
for $T\le T_{\rm F}$. If one assumes a functional form for
$\langle\sigma_a v\rangle_T$, an order-of-magnitude estimate of
$T_{\rm F}$ is given by the condition $n_X\langle\sigma
v\rangle_{T_{\rm F}}=H(T_{\rm F})$.  A more accurate formula is
obtained by a semi-analytical treatment of the Boltzmann
equation~\cite{ket89}
\begin{equation}\label{relicfreeze}
x_{\rm F}=x_0-(n+{\textstyle\frac{1}{2}})\,\log x_0\,,
\end{equation}
where
\begin{equation}
x_0=\log\left[0.038\,(n+1)\frac{g_X}{\sqrt{g_{*\rm F}}}m_{\rm Pl}m_X
\sigma_n\right]\,.
\end{equation}
Here, the parameterization
\begin{equation}\label{ratparam}
\langle\sigma_a v\rangle_T=\sigma_nx^{-n}
\end{equation}
with $x=m_X/T$ was assumed.  The most interesting cases are obtained
for $n=0$ (s-wave annihilation) and $n=1$ (p-wave). The previous
equations yield
\begin{eqnarray}
\left(\frac{n_X}{s}\right)_0&=&
\frac{3.79\,\sqrt{g_{*\rm F}}\,(n+1)\,x_{\rm F}^{n+1}}
{g_{s*\rm F} m_{\rm Pl}m_X \sigma_n}\,,\nonumber\\
\Omega_X h^2&=&
0.032\,\frac{(n+1)\,x_{\rm F}^{n+1}\sqrt{g_{*\rm F}}}
{g_{s*\rm F}(\sigma_n/{\rm pb})}\,, \label{relicomsig}
\end{eqnarray}
with an accuracy better than $5\%$. Since $\Omega_{\rm DM}h^2\approx
0.11$ \cite{Spergel:2003cb}, and $x_{\rm F}= 15$--20, it is easily
seen that, for $X\neq\bar{X}$, one has $\sigma_0\sim 5$ pb and
$\sigma_1\sim 10^2$ pb.

\section{Kinetic Freeze-Out}\label{kinfout}

We estimate the kinetic freeze-out temperature for $X$-particles
coupled to the electromagnetic plasma.  We make the rough
approximation that the annihilation and scattering cross sections are
comparable, $\sigma_{a} \sim \sigma_{s}$, so that the ratio between
the annihilation and scattering rates is essentially given by
$\Gamma_{a}/\Gamma_{s}\sim n_X/n_{s}$.  For our case the scattering
targets are electrons and positrons so that
\begin{equation}\label{neppnem}
n_{s} = n_{e^+}+n_{e^-} =
4\left(\frac{m_e^2}{2\pi z}\right)^{3/2}
e^{-z}\cosh\xi_e\,,
\end{equation}
where $\xi_e\equiv \mu_e/T$ is the electron degeneracy parameter and
$z \equiv m_e/T$.  Therefore, as long as the $X$-particles are in
kinetic equilibrium we have
\begin{equation}
\frac{\Gamma_{a}}{\Gamma_{s}}\approx
\frac{1}{2}\left(\frac{m_X}{m_e}\right)^{3/2}
\exp\left(-\frac{m_X-m_e}{T}\right)\frac{1}{2\cosh \xi_e}
\end{equation}
This result applies at $T_{\rm F}$ only if
$(\Gamma_{a}/\Gamma_{s})_{T_{\rm F}}\ll 1$.  Assuming $m_X=1$~MeV, one
finds from Eq.~(\ref{relicfreeze}) that $T_{\rm F}\approx0.07$~MeV and
from a standard BBN code that $\xi_e(T_{\rm F}) \approx 0.32\times
10^{-7}$ so that $(\Gamma_{a}/\Gamma_{s})_{T_{\rm F}} \approx 4\times
10^{-4}$ which is indeed ${}\ll1$.

Once a value for the annihilation effective cross section $\sigma_n$
has been determined, one can easily deduce a kinetic freeze-out
temperature $T_{\rm K}$ for the decoupling of the dark matter
particles from the electromagnetic plasma, assuming that $\sigma_a
\sim \sigma_s$. The condition $\Gamma_{s}(T_{\rm K})=H(T_{\rm K})$
implies
\begin{equation}\label{rateatTK}
\langle\sigma_{s}v\rangle_{T_{\rm K}}=
\frac{5.44\,T_{\rm K}^2}{m_{\rm Pl}n_{s}(T_{\rm K})}
\sqrt{\frac{g_{*}(T_{\rm K})}{10.75}}\,.
\end{equation}
With $\cosh\xi_e \approx 1$ and $g_{*}\approx 3.36$
one obtains
\begin{equation}\label{icskappa}
z_{\rm K}-(0.5-\ell)\log z_{\rm K}\approx 
14.1+\log(\sigma_{s,\ell}/{\rm pb})\,,
\end{equation}
where $\langle\sigma_{s}v\rangle\equiv\sigma_{s,\ell}z^{-\ell}$.  For
typical values of $\sigma_{s,n}$ this gives $T_{\rm K}\approx 35$~keV.

\section{Entropy generation during decoupling}     \label{decouplentr}

We sketch an argument that indicates that the entropy production
associated with the dark-matter freeze-out is a sub-leading effect.
To this end we assume that every $e^{+}e^{-}$ annihilation product is
instantaneously thermalized and converted into photons. In this case
the evolution equation for the $X$ number density is~\cite{ket89}
\begin{equation}
\dot{n}_X+3Hn_X=-C[n_X]
\end{equation}
where
\begin{equation}
C[n_X]=\langle\sigma_a v\rangle
\left[n_X^2-n_{X,{\rm eq}}^2\right]\,.
\end{equation}
The same equation applies to $n_{\bar X}=n_{X}$ because we always
assume equal distributions for both $X$ and $\bar X$.
The subsequent $e^+e^-$ annihilations imply
\begin{equation}
\dot{n}_\gamma+3Hn_\gamma=+2C[n_X]\,.
\end{equation}
The equilibrium energy densities are
\begin{eqnarray}
\rho_\gamma&=&
\left(\frac{\pi^2}{15}\right)
\left(\frac{\pi^2}{2\zeta(3)}\right)^{4/3}n_\gamma^{4/3}\,,
\nonumber\\
\rho_X&=&
n_X\left(m_X+\frac{3}{2}T\right)\,,
\end{eqnarray}
where the non-relativistic regime was used for the dark-matter
particles.
The energy injection by late annihilations is estimated as
\begin{eqnarray}
\dot{\rho}_\gamma&=&
\frac{8}{3}\left(\frac{\pi^4T}{30\zeta(3)}\right)C[n_X]\,,
\nonumber\\
\dot{\rho}_X&=&-m_X\bigg(1+\frac{3T}{2m_X}\bigg)C[n_X]\,.
\end{eqnarray}
The direct change of the total energy density due to this conversion
of dark-matter particles into photons is a very small effect, as one
can see by evaluating the ratios $\dot{\rho}_\gamma/\rho_\gamma$ and
$\dot{n}_\gamma/n_\gamma$.
 
The main non-negligible consequence is on the $T$-$t$ relation (first
Friedmann equation) through the second term in the following equation
\begin{eqnarray}
\frac{d\rho_{\rm em}}{dt}&=&-3H(\rho+P)_{\rm em}\nonumber\\
&&{}-\left[m_X-\left(\frac{4\pi^4}{45\zeta(3)} -3\right)T\right]
\,C[n_X]\,.
\end{eqnarray}
The problem is then restricted to the solution of the equation for
$n_X$ in order to calculate the relevant quantity $C[n_X]$.  This was
done by standard techniques after some standard substitutions (see
e.g.\ Ref.~\cite{ket89}).  We found a negligible effect on $Y_{\rm
p}$, and a change of the $^2$H and $^7$Li yields of order $0.1$ in the
units of Fig.~\ref{fig:emcoupled}.  Therefore, we are indeed dealing
with a sub-leading effect.

\section{Deuterium Photo-Dissociation}                 \label{Ddissoc}

We show that the dissociation of fragile nuclides, principally
deuterium, by late dark-matter annihilation is completely negligible.
To this end we first follow standard works on this subject
(e.g.~Ref.~\cite{Kawasaki:1994sc}) and note that the
``first-generation'' of nonthermal photons is rapidly degraded to have
a spectrum with a high energy cut $E_{\rm C}\approx m_e^2/22T$ because
of the highly efficient pair creation and subsequent reactions on the
background medium.  This means that in order for some photons to
survive and to be able to dissociate nuclei the temperature has to
drop at least to values $T\alt m_e^2/22B_{\rm D} \approx5$~keV, where
$B_{\rm D}\approx 2.2$~MeV is the binding energy of deuterium.  This
implies that neglecting this phenomenon during the BBN epoch is
certainly a good approximation.  Note also that in our prefered
models, MeV-mass $X$ particles would not annihilate directly into
photons, so the ``first-generation'' photon spectrum is already
degraded in energy because it is produced by secondary effects of the
primary electrons and positrons.

Therefore, by simply looking at the expected energy spectrum, the most
dangerous process would be the electro-disintegration of deuterium,
for which a further $0.5$~MeV penalty in the energy should be
considered as the electron rest mass would not be released.  Without
invoking a detailed analysis we conclude that our results up to
$m_X\alt 2.7$~MeV remain unchanged.

For higher values of $m_X$ considered in our BBN analysis, say
3--20~MeV, we reach the same conclusion by following the treatment of
the late-time annihilations of a relic particle described in
Ref.~\cite{Frieman:1989fx}.  Assuming that electro-dissociation of
deuterium is the only relevant phenomenon, the D depletion factor can
be written as
\begin{equation}
\exp\left[-\int_{t_i}^{t_f}dt\,\Gamma_{e{\rm D}}(t)\right]\,,
\label{eDdepletion}
\end{equation}
where $t_i$ is the time at the onset of dissociation, $t_f$ is
some late time where the D abundance is observed, and 
\begin{equation}
\Gamma_{e{\rm D}}=s\int_{B_{\rm D}}^{E_{\rm C}(T)}dE\,f_e(E,T)
\,\sigma_{e{\rm D}}(E)\,.
\end{equation}
Here, $\sigma_{e{\rm D}}(E)\approx 13.3~\mu{\rm b}\, [(E-B_{\rm
D})/{\rm MeV}]^{1/2}$ is the electro-dissociation cross-section of
deuterium~\cite{Picker84}.  Further, $f_e(E,T)$ is the ``steady
state'' dark-matter annihilation product spectrum.  By arguments
similar to the ones presented in Ref.~\cite{Frieman:1989fx} it is
written as
\begin{eqnarray}
f_e(E,T)&\approx&
\sigma_n\left(\frac{T}{m_X}\right)^n
\frac{s}{n_\gamma\sigma_{e\gamma}(E)}
\left(\frac{n_X}{s}\right)^2\nonumber\\
&&{}\times\frac{m_X\theta[E_{\rm max}(T)-E]}{\sqrt{E^3E_C(T)}}\,,
\label{ssdistr}
\end{eqnarray}
where $\sigma_{e\gamma}(E)$ is the Compton scattering cross section
and $E_{\rm max}(T)={\rm Min}[E_{\rm C}(T),m_X]$.  Note that the
primary particle's ``first'' shower spectrum has roughly the same
shape for $\gamma$ or $e$ initiated
showers~\cite{Frieman:1989fx,Kawasaki:1994sc}.

We have numerically solved the integral in Eq.~(\ref{eDdepletion}) by
changing to the temperature variable and assuming a radiation
dominated universe. The effects we found are completely negligible
both for s- and p-wave annihilations.

Note, however, that for showers induced by secondary photons one
should replace in Eq.~(\ref{ssdistr}) the huge factor $n_\gamma$ with
$n_e$, and the electro-dissociation cross section with the
photo-dissociation one, while a penalty factor in the energy spectrum
would enter. Making some extreme assumption one could thus gain up to
a factor $10^{12}$ of the previous estimates. Even so, it is not
enough to affect by more than a few percent our simplified nuclides'
predictions. The effects are even less pronounced for the preferred
case $n=1$ (p-wave annihilation).


\end{document}